%% file: bare_conf.tex
\documentclass[conference]{IEEEtran}
\usepackage[utf8]{inputenc}
\usepackage[newfloat]{minted}
\usepackage{quantikz}
\usepackage[colorlinks]{hyperref}
\usepackage{svg}
\usepackage{amsmath}
\usepackage{mathtools}

\usepackage{amssymb}% http://ctan.org/pkg/amssymb
\usepackage{pifont}% http://ctan.org/pkg/pifont
\usepackage{graphicx}
\usepackage{algpseudocode}
\usepackage{algorithm}
\usepackage[justification=centering]{caption}
\usepackage{subcaption}
\usepackage{placeins}

\usepackage{array}
\newcolumntype{P}[1]{>{\centering\arraybackslash}p{#1}}
\newcolumntype{M}[1]{>{\centering\arraybackslash}m{#1}}
\newcolumntype{Q}[1]{>{\arraybackslash}m{#1}}
\usepackage{comment}
\usepackage{float}
\usepackage{makecell}
\newcommand{\linebreakand}{%
  \end{@IEEEauthorhalign}
  \hfill\mbox{}\par
  \mbox{}\hfill\begin{@IEEEauthorhalign}
}

\usepackage{tikz}
\usetikzlibrary{shapes.geometric, arrows, decorations.pathmorphing}
\tikzstyle{startstop} = [rectangle, rounded corners,text centered, draw=black, fill=red!30]
\tikzstyle{io} = [trapezium, trapezium left angle=70, trapezium right angle=110, minimum width=3cm, minimum height=1cm, text centered, draw=black, fill=blue!30]
\tikzstyle{process} = [rectangle, text centered, draw=black, fill=orange!30]

\tikzstyle{decision} = [diamond, text centered, draw=black, fill=green!30]
\tikzstyle{arrow} = [thick,->,>=stealth]

\newcommand{\vew}[1]{
	\edef\start{\the\pgfmatrixcurrentrow-\the\pgfmatrixcurrentcolumn}
	\edef\end{\the\numexpr#1+\pgfmatrixcurrentrow\relax-\the\pgfmatrixcurrentcolumn}
	\expandafter\expandafter\expandafter\vewexplicit\expandafter\expandafter\expandafter{\expandafter\start\expandafter}\expandafter{\end}
}
\newcommand{\vewexplicit}[2]{
	\arrow[from=#1,to=#2,arrows,decorate,decoration={snake,amplitude=1pt,segment length=6.5pt}] {}
}
\usepackage{listings}
\usepackage{xcolor}

\definecolor{codegreen}{rgb}{0,0.6,0}
\definecolor{codegray}{rgb}{0.5,0.5,0.5}
\definecolor{codepurple}{rgb}{0.58,0,0.82}
\definecolor{backcolour}{rgb}{0.95,0.95,0.92}

\lstdefinestyle{mystyle}{
    backgroundcolor=\color{backcolour},   
    commentstyle=\color{codegreen},
    keywordstyle=\color{magenta},
    numberstyle=\tiny\color{codegray},
    stringstyle=\color{codepurple},
    basicstyle=\ttfamily\footnotesize,
    breakatwhitespace=false,         
    breaklines=true,                 
    captionpos=b,                    
    keepspaces=true,                 
    numbers=left,                    
    numbersep=5pt,                  
    showspaces=false,                
    showstringspaces=false,
    showtabs=false,                  
    tabsize=2
}

\usepackage{adjustbox}
\usepackage{multirow}
\usepackage{amssymb}% http://ctan.org/pkg/amssymb
\usepackage{pifont}% http://ctan.org/pkg/pifont
\usepackage{siunitx}

\begin{document}
\bstctlcite{IEEEexample:BSTcontrol}

% \title{Compiler for diamond Based quantum computers}
% \title{Hardware-aware compiler tailored for diamond NV centers}
% \title{Exploiting hardware specific characteristics for controlling diamond NV centers, using a compiler}
\title{Compiler design for hardware specific decomposition optimizations, tailored to diamond NV centers}

\author{\IEEEauthorblockN{Folkert de Ronde} 
    \IEEEauthorblockA{Quantum \& Computer Engineering \\
        % MSc Quantum Electrical Engineering \\
        Delft University of Technology \\
        Delft, The Netherlands \\
        f.w.m.deronde@tudelft.nl}
\and 
\IEEEauthorblockN{Stephan Wong}
\IEEEauthorblockA{
Quantum \& Computer Engineering \\
Delft University of Technology \\
Delft, The Netherlands \\
j.s.s.m.wong@tudelft.nl}
\and 
% \linebreakand
\IEEEauthorblockN{Sebastian Feld}
\IEEEauthorblockA{
Quantum \& Computer Engineering \\
Delft University of Technology \\
Delft, The Netherlands \\
s.feld@tudelft.nl}}
% \and
% \IEEEauthorblockN{David Elkouss}
% \IEEEauthorblockA{Networked Quantum Devices Unit, Okinawa Institute of Science and Technology Graduate University, Okinawa, Japan\\
% QuTech, Delft University of Technology, Delft, The Netherlands \\
% david.elkouss@oist.jp}}
%\and
%\IEEEauthorblockN{James Kirk\\ and Montgomery Scott}
%\IEEEauthorblockA{Starfleet Academy\\
%San Francisco, California 96678--2391\\
%Telephone: (800) 555--1212\\
%Fax: (888) 555--1212}}

\maketitle
% \newpage
\thispagestyle{plain}

\input{Chapters/abstract}

\input{Chapters/introduction}

\input{Chapters/related_work}

\input{Chapters/methodology}
\input{Chapters/results}
% \input{Chapters/discussion}
\input{Chapters/Conclusion}
\input{Chapters/acknowledgement}
% \clearpage
\bibliographystyle{IEEEtran}
\bibliography{report.bib}
% \printbibliography

% \bibliography{references.bib}

% \newpage

\end{document}

%% file: Chapters/abstract.tex
\begin{abstract}

Advances in quantum algorithms as well as in control hardware designs are continuously being made. These quantum algorithms, expressed as quantum circuits, need to be translated to a set of instructions from a defined quantum instruction-set architecture (ISA), which are executed by the control hardware. These translations can be done by a compiler, targeting different qubit technologies. Specifically for diamond NV centers, no compiler exists to perform this translation.  Therefore, in this paper we present a compiler designed for quantum computers utilizing diamond NV center specific instructions, such as direct carbon control and partial swaps, to reduce execution times and gate count. Additionally, our compiler adds on top of general compilers by allowing classical instructions to perform state tomography and measurement-based operations. 
The output of the compiler is tested in a diamond NV center specific simulator. Comparing a general compiler output with the diamond NV center specific output of our compiler while applying decoherence and depolarization noise showed reduced noise effects due to diamond specific decomposition. The compiler was also tested to perform state tomography and measurement-based operations, which showed to be functional. 
Our results show that we have successfully created a compiler with integrated classical and quantum instructions support, which can improve circuit execution fidelity by utilizing diamond specific optimizations.

\end{abstract}

%% file: Chapters/introduction.tex
\section{Introduction}\label{sec:intro}

% Talk about the other compilers that do not use cx carbon control, qubit typing, parameters for operation control, half swaps, diamond specific control operations (crc, entangle) that do not have support for classical instructions, or support for gate cancellation between swaps and performing operations on best suitable candidate.

% we are new due to the fact that we combine classical with quantum, in a manner which is specific to diamond nv centers.

% This is also where we give the background information about the systems. Why do we need something that performs all these specific operations for diamond based nv centers?
% In the latest years of quantum computing, advances are made in creating new algorithms and improving the underlying hardware needed to perform quantum operations. However, there is a critical component missing, which is the translation of quantum algorithms into executable instructions suitable for hardware implementation. This translation can be done by the use of a compiler, which must have knowledge about the underlying hardware system, including the type and parameters of qubits in the system.

Recent advancements in quantum computing have led to the development of new algorithms and improvements in the underlying hardware necessary for executing quantum operations. Such hardware are often based on the execution of a (newly defined) instruction-set architecture (ISA), which enables the utilization of compilers. % In order to execute these quantum algorithms, they need to be translated into instructions executable by hardware. Next to this, we are currently in the NISQ era \cite{NISQ}, where the fidelity of the execution of quantum circuits on hardware is heavily influenced by noise. In order to reduce noise effects in a quantum circuit, the amount of operations performed and total execution time can be reduced. 

Translating a quantum circuit into ISA compatible instructions can be performed by a compiler, \cite{CompilerTranspilation}, which is now commonly solved by using a generic compiler \cite{rattacaso2024quantum} or a compiler created for a specific set of hardware (such as spin qubits \cite{SpinQ}). The goal of such a compiler is to create a set of instructions that is capable of executing a quantum circuit on hardware. Since different quantum hardware architectures exist -- such as superconducting qubits, spin qubits, and diamond color center qubits -- a specialized compiler can leverage detailed knowledge of the underlying system to optimize performance.

Moreover, we are currently in the Noisy Intermediate-Scale Quantum (NISQ) era \cite{NISQ}, and the fidelity of quantum circuit execution is significantly affected by noise. One way to mitigate noise is by reducing the number of operations and the overall execution time of a quantum circuit.

\begin{figure}
    \centering
    \includegraphics[width=0.8\linewidth]{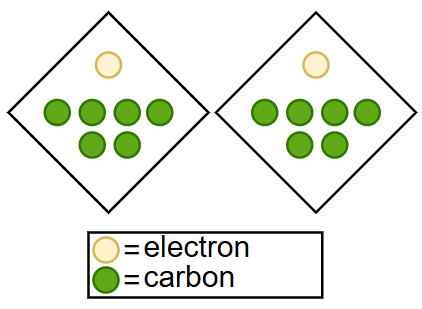}
    \caption{\textit{A visual representation of the qubits within a diamond NV center. Presenting both the electron and carbon qubits in the system. In this system 2 qubit gates are possible from electron on carbon and entanglement generation is possible between electrons on different NV centers.} }
    \label{fig:diamond_representation}
\end{figure}

For instance diamond color centers make use of multiple types of qubits, electron and carbon qubits, as depicted in Figure \ref{fig:diamond_representation}. The carbon qubits are connected to the electron via star-graph connectivity, making the electron the communication qubit of the system. In Figure \ref{fig:diamond_representation} ,two diamond Nitrogen-Vacancy  (NV) center nodes are presented with one electron (yellow) and multiple carbon qubits (green). These qubits require distinct control mechanisms, and a compiler designed specifically for NV centers can exploit system-specific optimizations such as smart routing and decomposition. These optimizations can result in reduced execution times and mitigate noise effects. Despite these advantages, such a compiler suited for diamond NV centers does not yet exist.

% We are in the nisq era and we have issues with performing quantum algorihtms because they take too long. We need to reduce the execution time, which we can do by using our diamond based optimizations.

Our research focuses on compiling circuit representations of quantum algorithms, described in Qiskit \cite{qiskit} or in Qassembly resembling OpenQASM \cite{IBM_architecture}, into instructions specifically designed for diamond NV center-based quantum systems. By catering to this particular quantum system, we can leverage diamond-specific optimizations, such as partial-swaps \cite{swapgates} and direct control \cite{swapgates} of carbon qubits.%$, resulting in lower execution times and less operations.

This topic holds significant interest for the broader scientific community for several reasons. Firstly, effective compilation is crucial for the precise control of quantum hardware, resulting in the potential to execute algorithms on the target hardware. Our compiler enables the automated conversion of any quantum circuit into executable instructions tailored for diamond systems.
Secondly, by utilizing diamond-specific optimizations, we aim to reduce execution times and mitigate noise effects associated with quantum operations, improving the decomposed circuit fidelity.

To achieve comprehensive control over the quantum system, the compiler can improve its output if it can incorporate knowledge of technology-specific operations.
% In our case, we focus on diamond NV centers\cite{NVcenter}, which can make use of specific techniques such as partial-swaps and direct carbon control\cite{swapgates}. 
The NV centers, next to the aforementioned partial-swaps and direct carbon control, also possess unique methods for qubit initialization (because carbon qubits cannot be initialized and measured directly, routing is needed) \cite{swapgates,beukers2024control}, system diagnostics \cite{CRC,Larmor,Larmor2,LarmorCarbon}, and entanglement generation \cite{entanglement}. Furthermore, the compiler must be capable of handling classical instructions to facilitate automatic execution of tasks, such as state tomography \cite{state_tomography} and classically controlled quantum operations (due to measurements). By integrating both quantum and classical control, the compiler has full control of the quantum hardware.
 
In this paper, we propose and demonstrate a new hardware-aware compiler that exploits specific hardware characteristics to control NV centers effectively, thereby reducing execution times for quantum programs. 
The compiler utilizes multiple techniques, based on the diamond specific operations, to create executable instructions for the target hardware. 
The foundation of the compiler is built in the Qiskit framework. This framework was chosen because it supports transpilation, single-qubit optimization techniques, and routing based on qubit connectivity. These methods are utilized by the compiler.
% The techniques are based on the previously mentioned  Multiple techniques are needed for the different functionalities of the compiler, such as partial swaps, direct control and classical support. 
The output of the compiler is simulated to verify the correct functionality of the compiler, and to test the effects of noise.

The research question addressed in this paper is: 
% How can a compiler leverage diamond specific operations to improve execution of quantum circuits in the target hardware.
How can a quantum compiler be adapted for diamond-based quantum computers to improve execution of quantum circuits?
 
The main contribution of this paper is a compiler that has the following high-level functionalities:
\begin{enumerate}
    \item \textbf{Classical Instruction Control:} 
    % Classical instructions are needed to control quantum hardware at multiple moments. Performing a classically controlled operation on a qubit based on a measurement value can only be done with classical control. Repeating a quantum algorithm many times to perform state tomography \cite{state_tomography} is much easier if the compiler creates instructions that perform this for you.
    % Our compiler can decompose classically controlled operation and state tomography into assembly language, which can be executed by the target hardware.
    
    Enables decomposition of classically controlled operations such as state tomography into executable assembly instructions for RISC-V based ISAs.

    \item \textbf{Enhanced Quantum Control:} 
    % In diamond NV centers special operations can be performed that reduce the total execution time and sometimes also the amount of operations that are needed. Our compiler can determine when to leverage these special operations to reduce operations and reduce execution time of quantum circuits.

     Identifies opportunities to use NV center-specific operations to reduce execution time and minimize gate count.

    \item \textbf{Qubit-Specific Instruction Execution:} 
    % The NV center makes use of multiple types of qubits, namely electrons and carbons. Both of these types of qubits require to be operated on in a different manner. For instance, the carbon qubit cannot directly be initialized or measured. Our compiler is capable of decomposing instructions dependent on the type of qubit that is operated on.

     Adapts operations based on qubit type (electron or carbon), ensuring proper initialization and measurement procedures.

    \item \textbf{Diamond NV Center Control:}
    % The Diamond NV center has a special instruction that performs an entanglement \cite{entanglement} between two electrons on distant NV centers.
    % Our compiler allows this special instruction to be supported. 
    
    Supports specialized entanglement instructions \cite{entanglement} between distant NV center electrons.
    \item \textbf{System Diagnostics:} 
    % Diamond NV centers are commonly operated on in a certain charge state \cite{CRC}. In order to ensure that the NV center is in the charge state, a charge resonance check needs to be performed. Other system diagnostics such as determining the frequency of the electron and carbon qubits need to be performed as well. 
    % Our compiler can automatically add these system diagnostics to the set of instructions. 

     Automates the addition of charge resonance checks and frequency calibrations to maintain optimal NV center operation \cite{CRC}.

\end{enumerate}

In Section \ref{sec:related_work}, we compare our work with existing compilers, we also highlight the gap between compilers created for specific hardwares and compilers created for diamond NV centers. In Section \ref{sec:Compiler_design}, we present the design of our compiler, accompanied by the needed background information, for diamond color centers, explaining the settings that contribute to its functionality. Section \ref{Sec:results_discussion} presents our results, demonstrating the compiler’s capabilitie to generate hardware executable instructions in an improved manner resulting in lower noise effects with regards to simple decomposition. We demonstrate that the compiler is capable of creating instructions implementing functional measurement-based operations and a sample algorithm. We also present our findings and discuss the limitations. Finally, Section \ref{sec:conclusion} offers conclusions regarding our findings and outlines potential future research.

%% file: Chapters/related_work.tex
\section{Related work} \label{sec:related_work}
Previous works have developed quantum compilers for various hardware platforms, such as neutral atoms \cite{neutral_atoms}, superconducting qubits \cite{superconducting} or spin qubits \cite{SpinQ}. However, these efforts have not specifically addressed diamond NV centers or leveraged hardware-specific operations to optimize quantum execution.

In \cite{Distributed_quantum_computing}, the focus is on distributed quantum systems, which include diamond color centers. However, the work emphasizes logical operations between distributed quantum systems rather than local optimizations and control.

In \cite{IBM_architecture}, the emphasis is on general quantum compilation, facilitating high-level, hardware-agnostic optimizations. Users can input backend details to generate OPENQASM 3.0 output with backend-compatible instructions. However, these instructions are limited to general operations executable on any hardware, lacking support for system-specific optimizations.

In the previous works, some problems are not addressed, such as leveraging hardware specific operations, local optimizations and system specific optimizations. The previous works lack focus on specific system implementation. In our case, we can make use of the limitations of the diamond NV center system and decompose our instructions in an improved manner, resulting in lower execution times and/or operations. These improvements result in higher execution fidelity of a quantum circuit. 

%% file: Chapters/methodology.tex
\section{Methodology} \label{sec:Compiler_design}
% /****

% how did we actually implement the new operations. 

% In this section we describe the different functionalities that are needed in order to have control of a diamond based quantum computer, as well as some functionalities that improve the control. 

% The functionalities that need to be added are:

% \begin{itemize}
%     \item Two qubit gate control of carbon qubits
%     \item Photon entanglement instructions
%     \item Full control of diamond ISA
%     \item Control of classical instructions
% \end{itemize}

% For multiple of these functionalities, it is important that it is known what type of qubit we are acting upon. For this reason we have developed a qubit typing algorithm, which casts a type to the qubit within the code module. 

% ****$\backslash$

% setup of this section
% - first introduction
% - main functionalities
% - introduction of global information
% - actual main functionalitie explanation

% This section presents our compiler design tailored for diamond based quantum computers. Quantum algorithms are generally not written for a certain quantum system. This means that a compiler is needed that translates a quantum algorithm into instructions that can be executed on quantum hardware. This quantum hardware has an instruction set architecture (ISA), to which the compiler must obhold.

This section presents our compiler designed specifically for diamond-based quantum systems, utilizing NV centers. Quantum algorithms are generally hardware-agnostic, requiring a compiler that translates their circuit representation into hardware-specific instructions. The target quantum hardware has its own Instruction Set Architecture (ISA), defining the set of executable instructions for both classical and quantum operations. In this work, we use the ISA proposed in \cite{ISA_base}, where the classical ISA is based on RISC-V \cite{riscv2019}.

Our primary goal is to develop a compiler with the high-level functionalities described in Section \ref{sec:intro}. These high-level functionalities are implemented by the following key functionalities:
\begin{enumerate}
    \item \textbf{Transpile instructions into diamond NV center specific instructions:} \\
    Ensures that the final quantum circuit consists only of instructions executable on NV center hardware. We do not yet aim for optimization at this stage.
    \item \textbf{Integrate classical and quantum decompositions:} \\
    Enables state tomography and measurement-based corrections by incorporating classical instructions where necessary. For example, classical feedback is essential for conditional operations in quantum algorithms.
    \item \textbf{Leverage diamond-specific instructions to mitigate noise effects:} \\
    Utilizes diamond-specific optimizations to minimize gate count and improve fidelity.
    \item \textbf{Optimize operation routing, considering the constraints imposed by the connectivity of the carbon atoms in the NV center:}\\
    Since only the electron qubit can be directly initialized and measured, operations on carbon qubits must be strategically routed.
\end{enumerate}

\begin{figure*}
    \centering
    \includegraphics[width=\linewidth]{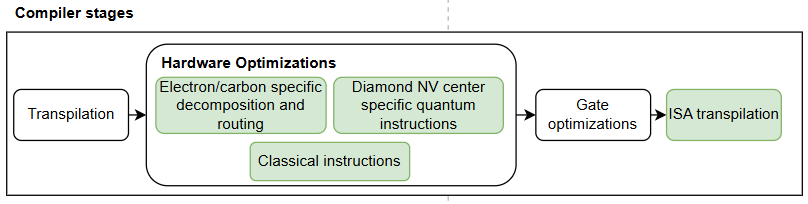}
    \caption{\textit{A representation of compiler stages, indicating the functionalities of the compiler. The stages presented in green are added by us, while the stages in white are already present in Qiskit.}}
    \label{fig:compiler}
\end{figure*}

These functionalities are achieved through four main compilation stages, illustrated in Figure \ref{fig:compiler}. The functionalities are implemented in the compiler stages indicated by the number of the functionalities. The stages indicated in green are fully created by us, while the other stages were already part of Qiskit. The hardware-aware compiler stage uses three passes to perform hardware-specific optimizations. These three functionalities are visualized in Fig \ref{fig:compiler} and presented in Sections \ref{sec:qubit_typing},\ref{sec:Quantum_instructions}, \ref{sec:classical_instructions}. 

\subsection{Electron/carbon specific decomposition and routing} \label{sec:qubit_typing}

Diamond NV centers host multiple types of qubits, primarily electrons and carbon nuclei, each requiring distinct control methods and instructions. Since different qubit types have varying ways to be controlled, the electron and carbon atoms for instance require different execution times and frequencies to perform the same rotation \cite{swapgates}. Therefore, the compiler must accurately identify and manage the type of qubit involved in each operation. 
Additionally, electrons can freely be controlled in diamond NV centers, while carbon atoms can only be controlled using direct gates where the electron is the control qubit \cite{swapgates,unden2019revealing,taminiau2014universal}. However, in order to execute an arbitrary quantum algorithm on the hardware, arbitrary rotation gates need to be performed on the carbon qubits. In order to achieve this, routing and special decomposittions are needed. 
% In diamond NV centers, electron and carbon qubits are operated in a different manner. Only free control of the electron is possible, while control over the carbon qubits need to be done while making use of the electron. Therefore, to execute instructions accurately, the compiler must identify the qubit type (electron or carbon).

\begin{figure}[]
    \centering
    \includegraphics[width=\linewidth]{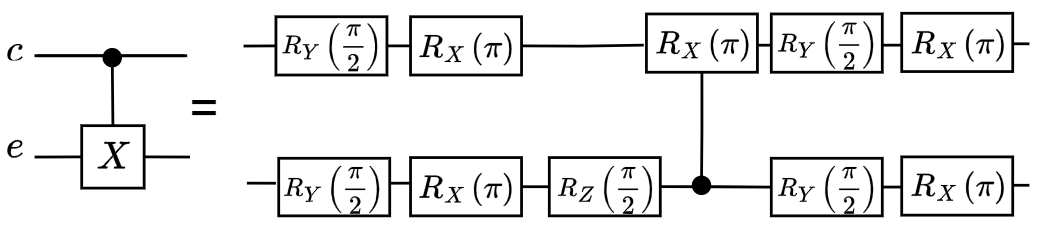}
    \caption{\textit{Quantum circuit depicting the decomposition of a cx gate from a carbon onto the electron in the native gateset of diamond NV centers. The left hand side of the figure shows the instruction to be executed, while the right hand side shows what operations are needed to realise the instruction in diamond NV centers.}}
    \label{fig:cx_flip_decomp}
\end{figure}

The different types of decomposition needed for carbon and electron qubits requires the compiler to be able to identify the qubit type and perform decompositions accordingly. 
Qubit identification ensures precise instruction routing, especially since carbon qubits cannot be directly initialized, measured or controlled. Instead, the electron qubit must mediate these operations:
\begin{itemize}
    \item \textbf{Initialization}: First, initialize the electron, then swap its state to the carbon qubit.
    \item \textbf{Controlled Gates}: Controlled operations can only be performed with the electron as the control qubit. If a controlled operation is needed between two carbon qubits, the information stored in the control carbon qubit must be swapped onto the electron. For control operations from carbon to electron, gate decomposition can reverse the control direction, as illustrated in Figure \ref{fig:cx_flip_decomp}.
    \item \textbf{Measurement}: Swap the carbon qubit state onto the electron, then perform the measurement.
\end{itemize}

Identifying which qubit is being operated on can be achieved by using some information about the system. In our system, the electron is always set to qubit 0 and the total number of qubits per NV center can be set dynamically, where the number of qubits per NV center is equal for every NV center. Using this information, the qubit type can be identified using Equation \ref{eq:qubit_type}, where any operation that is acted on multiple qubits is always a carbon operation, given that 2-qubit operations can only be directed from electron to carbon.
\begin{equation}
    q_{i_{type}} = \begin{cases} \mbox{'carbon'} & \mbox{if } q_{i} \mbox{ mod } \#qubits\_per\_NV \neq 0\\ & \mbox{or multi\_qubit\_operation} \\
    \mbox{'electron'} & \mbox{otherwise}
    \end{cases}
    \label{eq:qubit_type}
\end{equation}

Using the qubit type, the compiler can decompose the instructions accordingly. When an operation is determined to be an electron operation, the compiler will create an ISA instruction that resembles the operation for an electron. Given that the hardware requires to know if the instruction is an electron or carbon instruction, the compiler will choose the instruction corresponding to an electron rotation (\textit{qgatee} \cite{ISA_base}).

% ?? sebastian, talk about the list of instructions

In the case that the qubit type of the operation is determined to be a carbon instruction, the compiler will determine what to do based on the instruction. For an initialization or a measurement operation, the qubit state will be swapped to the electron and the instruction will be performed. In the case of a controlled gate, the solution is less trivial and will be explained in Section \ref{sec:Quantum_instructions}

\subsection{Diamond NV center specific quantum instructions}\label{sec:Quantum_instructions}

Given the connectivity limitations inherent in Diamond NV centers, various optimization techniques can be employed to enhance performance. One notable approach is direct control of carbon qubits \cite{swapgates}, which significantly reduces execution time compared to electron-mediated control. To harness these hardware-specific optimizations, both the compiler and the hardware must be designed to support these specialized instructions. While electron-mediated control of carbon qubits remains common, direct control methods offer a faster alternative under certain conditions.

Diamond-specific quantum instructions can be put into two categories. The instructions needed to perform system diagnostics, and the instructions that are used to reduce the noise effects during execution of an algorithm on hardware. 

\subsubsection{System diagnostics}
Before executing a quantum circuit, system diagnostics must be performed to ensure the NV center operates reliably. In diamond NV centers, these diagnostics involve determining the Larmor frequency and Rabi frequencies of both electron and carbon qubits. Additionally, the Charge Resonance Check (CRC) \cite{CRC} ensures the NV center is in the correct negative charge state \cite{ChargePump}. The CRC must be conducted before the algorithm starts and repeated afterwards to verify that the NV center did not ionize during execution. If the system did ionize, the execution of the circuit has failed and the results do not actually hold any usefull information regarding the executed algorithm in them. Therefore, the results should be discarded and the system should be put back in the correct charge state, for instance by making use of a charge pump \cite{ChargePump}.

The CRC cannot be performed in the middle of a circuit, because the CRC will destroy all the information stored in the qubits on the system. The compiler adds the specified system diagnostics at the start of every algorithm and adds the CRC at the end of every quantum algorithm as well. The sequence of diagnostic tests is crucial due to dependencies between tests:
\begin{enumerate}
    \item The electron Larmor frequency test must precede the Rabi test, because the frequency needs to be known in order to rotate the electron.
    \item Electron diagnostics must precede carbon diagnostics, as carbon qubit measurements rely on the electron’s Larmor and Rabi frequencies.
\end{enumerate}

The resulting diagnostic sequence is as follows:

\begin{lstlisting} 
LarmorElectron 
RabiCheckElectron 
LarmorCarbon 
RabiCheckCarbon 
CRC 
QuantumAlgorithm 
CRC 
\end{lstlisting}
% ?? why show as dependancy graph?

This structured diagnostic approach ensures the NV center is correctly initialized and stable throughout the algorithm's execution, minimizing the risk of errors caused by ionization.

\subsubsection{Improved operation decompositions}
Two types of instructions can be optimized for decomposition when using diamond NV centers: single qubit carbon rotations and swaps \cite{swapgates}. 

\begin{figure}[H]
    \centering
    \includegraphics[width=0.7\linewidth]{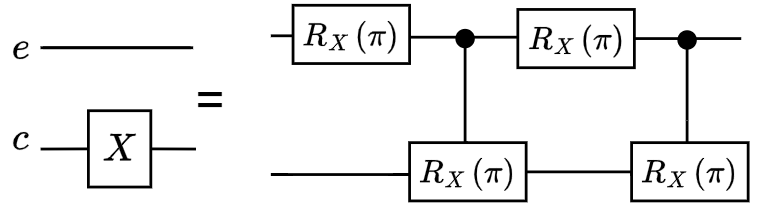}
    \caption{\textit{Quantum circuit representing an single qubit rotation gate on the carbon when the electron state needs to be preserved (DDrf gate). The left hand side shows the operation that needs to be executed, while the right hand side shows a solution to perform the instruction on diamond NV centers while preserving the state of the electron.}}
    \label{fig:no_rotation_gate}
\end{figure}

\begin{figure}[H]
    \centering
    \includegraphics[width=0.65\linewidth]{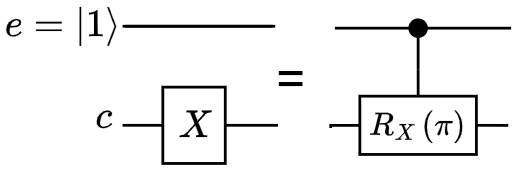}
    \caption{\textit{Quantum circuit representing a single qubit rotation gate on the carbon when the electron state does not need to be preserved (direct control). The left hand side shows the operation that needs to be executed, while the right hand side shows a solution to perform the instruction on diamond NV centers while destroying the state of the electron, but reducing the amount of instructions needed.}}
    \label{fig:rotation_gate}
\end{figure}

To perform single qubit operations on a carbon qubit without altering the electron’s state, a sequence called the DDrf \cite{swapgates} gate is required. Figure \ref{fig:no_rotation_gate} illustrates a simplified version of the DDrf gate, in which the electron’s state is flipped between operations to eventually preserve the electron state. However, when preserving the electron state is not needed, the electron can be initialized directly into the $\ket{1}$ state, allowing direct control on the carbon qubit. This approach, shown in Figure \ref{fig:rotation_gate}, is more efficient, taking half the time and reducing the risk of introducing noise. The reason why this halves the time, is because electron operations are much faster then carbon operations, in the order of 100ns vs. 1-2 ms. This means that the reduction of one carbon gate is the most important reduction in this optimization. 

The compiler must thus be able to determine if the electron state needs to be preserved. The electron qubit state is deemed irrelevant in the following scenarios:
\begin{enumerate}
    \item No prior operations have been performed on the electron since the start of the algorithm.
    \item No operations have occurred on the electron since its last measurement.
    \item The electron qubit will never be operated on before the end of the algorithm (e.g., when its state is swapped onto a carbon qubit for storage).
    \item The next instruction supposed to happen on the electron is an initialization instruction.
\end{enumerate}
% ?? point 3 above, what if it is used as control. Lets talk about this sebastian

By identifying these conditions, the system can determine whether to employ a DDrf gate or direct control, ensuring optimal performance while minimizing unnecessary operations.

\begin{figure}
    \centering
    \includegraphics[width=0.63\linewidth]{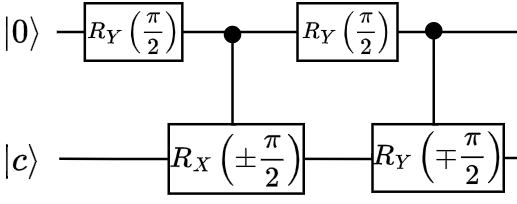}
    \caption{\textit{Quantum circuit used to perform initialization of a carbon qubit. \cite{swapgates}}}
\label{circ:initialization}
\end{figure}

\begin{figure}
    \centering
    \includegraphics[width=0.6\linewidth]{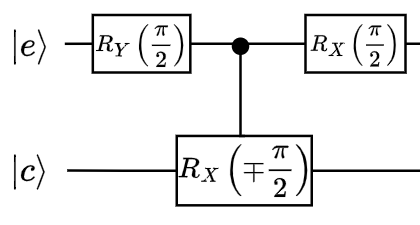}
\caption{\textit{Quantum circuit used to swap the carbon qubit state to the electron qubit in the X basis. \cite{swapgates}}}
  \label{circ:meas_carbon_X}
\end{figure}

\begin{figure}
    \centering
    \includegraphics[width=0.6\linewidth]{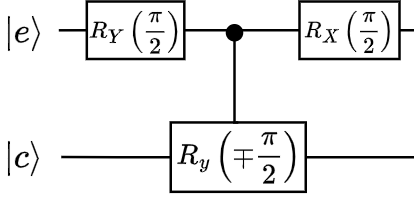}
\caption{\textit{Quantum circuit used to swap the carbon qubit state to the electron qubit in the Y basis. \cite{swapgates}}}
  \label{circ:meas_carbon_Y}
\end{figure}

\begin{figure}
    \centering
    \includegraphics[width=0.62\linewidth]{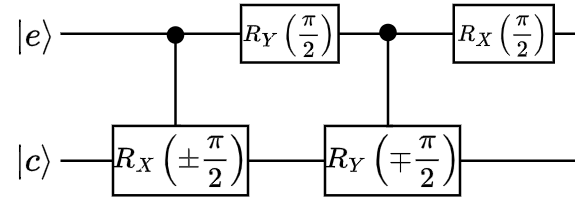}
\caption{\textit{Quantum circuit used to swap the carbon qubit state to the electron qubit in the Z basis. \cite{swapgates}}}
  \label{circ:meas_carbon_Z}
\end{figure}

The second type of optimized instructions in diamond NV centers involves swap operations, specifically partial-swaps. Partial-swaps are crucial for performing measurements in specific bases and for qubit initialization\cite{swapgates}. By leveraging partial swaps, the number of required instructions for performing intialization and measurement can be reduced, optimizing the execution of the quantum algorithm. These swaps can be performed in various bases -- X, Y, and Z -- as illustrated in Figures \ref{circ:meas_carbon_X}, \ref{circ:meas_carbon_Y}, and \ref{circ:meas_carbon_Z}, respectively. In the figures the $\pm$ signs mean that the direction of the controlled gate is $+$ in the $0$ case and $-$ in the $1$ case, while the $\mp$ signs mean the opposite.  These partial swaps swap the qubit state of a carbon qubit on the electron qubit, but all information that was stored in the electron is lost. The benefit here is that the swap requires fewer operations, while the downside is that the electron state loses stored information.  The swap, depicted in Figure \ref{circ:initialization}, can be used to initialize a carbon qubit in the $\ket{0}$ state, again using less operations, but the information stored in the state of the electron qubit is again lost. 

The compiler can determine whether a full swap or partial swap is necessary using the same logic employed for direct carbon control. If the electron’s state is deemed irrelevant, it is used to facilitate a partial swap. Next to this the compiler needs to be able to detect which type of swap needs to be performed, given the basis. Any measurement is presented in the Z basis, so identifying which basis to perform the swap in can be achieved by use of the following logic:
\begin{enumerate}
    \item If the predecessor of the measurement operation (on the carbon) is a Hadamard gate, the measurement is interpreted as occurring in the X basis, and the X-basis swap is applied.
    \item If the predecessors include a phase gate followed by a Hadamard gate, the measurement is classified as occurring in the Y basis, prompting a Y-basis swap.
    \item If none of the previously mentioned operations precede the measurement, the swap defaults to the Z basis.
\end{enumerate}
By dynamically selecting the swap basis, the system ensures efficient and accurate measurements, reducing unnecessary operations and enhancing the overall performance of the quantum algorithm.

% \begin{figure}[H]
% \begin{center}
%     \begin{quantikz}
%     \ket{e} &  \gate{R_y(\pi/2)}  & \ctrl{1} & \gate{R_x(\pi/2)} \\
%     \ket{c} &  \qw & \gate{R_x(\mp \pi/2)} & \qw
%     \end{quantikz}
% \end{center}
% \caption{Quantum circuit used to swap the carbon qubit state to the electron qubit in the X basis. \cite{swapgates}}
%   \label{circ:meas_carbon_X}
% \end{figure}

% \begin{figure}[H]
% \begin{center}
%     \begin{quantikz}
%     \ket{e} &  \gate{R_y(\pi/2)}  & \ctrl{1} & \gate{R_x(\pi/2)} \\
%     \ket{c} &  \qw & \gate{R_y(\mp \pi/2)} & \qw
%     \end{quantikz}
% \end{center}
% \caption{Quantum circuit used to swap the carbon qubit state to the electron qubit in the Y basis. \cite{swapgates}}
%   \label{circ:meas_carbon_Y}
% \end{figure}

% \begin{figure}[H]
% \begin{center}
%     \begin{quantikz}
%     \ket{e} & \ctrl{1} &  \gate{R_y(\pi/2)}  & \ctrl{1} & \gate{R_x(\pi/2)} \\
%     \ket{c} &  \gate{R_x(\pm \pi/2)} & \qw & \gate{\mp R_y(\pi/2)} & \qw
%     \end{quantikz}
% \end{center}
% \caption{Quantum circuit used to swap the carbon qubit state to the electron qubit in the Z basis. \cite{swapgates}}
%   \label{circ:meas_carbon_Z}
% \end{figure}

\subsection{Classical instructions} \label{sec:classical_instructions}
Effective quantum computation in diamond NV centers requires integration of classical and quantum instructions. For example, classical instructions are essential for iterative operations, such as state tomography \cite{state_tomography}, where repeated measurements are needed to reconstruct quantum states. Furthermore, classical logic is critical for enabling measurement-based operations \cite{Roffe03072019} and data storage \cite{classicalIntegration}, allowing dynamic branching based on quantum measurement outcomes. This capability ensures robust support for complex quantum algorithms that depend on conditional execution and feedback control, examples of such algorithms are quantum error correction codes (QECCs) \cite{stabilizer_surface_codes}. 
% Classical registers are employed to store intermediate results, control loop execution, and manage branching based on measurement outcomes.

% A quantum computer operates through a combination of classical and quantum instructions. This integration is essential for tasks such as repeated operations (such as state tomography \cite{state_tomography}), data storage \cite{classicalIntegration}, and measurement-based operations REF?. 

\subsubsection{State Tomography}
To perform state tomography \cite{state_tomography}, the hardware must run multiple iterations of the quantum circuit. The compiler can automatically create a set of instructions that result in the execution of a quantum algorithm multiple times, as presented in Listing \ref{lst:state_tomography}. The compiler generates instructions to measure the relevant data qubits (user defined) in each iteration and store the results in memory. The number of iterations and the relevant qubits to be measured can be configured by the user (identified in the listing by ``UserDefined'' and ``UserDefinedQubitx''), ensuring flexibility and precision in data collection.

\begin{lstlisting}[caption={State tomography example},label={lst:state_tomography}]
    LDi 0 RepetitionCounter
    LDi "UserDefined" RepetitionAmount
    label Repeat
    DecomposeQuantumCircuit(circuit)
    "Decomposition of quantum Circuit"
    Measure "UserDefinedQubit0"
    Measure "UserDefinedQubit1"
    ST MeasureResultRegister0
    ST MeasureResultsRegister1
    ADDi RepetitionCounter 1
    BR RepetitionCounter < RepetetionAmount Repeat
\end{lstlisting}
% ?? not that important subsec, lets talk
\subsubsection{Measurement-based operation}
When a classically controlled operation needs to be executed, the compiler generates a branching instruction that skips specific operations based on the measurement value stored in the measurement register. This conditional branching allows efficient control flow based on quantum outcomes.

An example of such a branching operation is shown in Listing \ref{lst:meas_based}.

\begin{lstlisting}[caption={Measurement-based X operation},label={lst:meas_based}]
    BR MeasurementRegister < 0 skip0
    Xgate qubit0
    label skip0
\end{lstlisting}
In this example, if the value in the MeasurementRegister is less than 0 (representing -1), the program jumps to the label skip0, bypassing the X gate operation on qubit0. This mechanism ensures handling of conditional quantum operations within the overall algorithm flow.

%% file: Chapters/results.tex
\section{Evaluation and discussion} \label{Sec:results_discussion}
To validate the functionality of the compiler, we evaluate its output by simulating quantum execution on a diamond NV center-based microachitecture simulator \cite{ISA_base}. The following sections present the results of different compiler tests, including validation of circuit decomposition, classically controlled operations, and improved quantum circuit execution fidelity by reduction of noise.
% The compiler tests include a test for validation of algorithm decomposition, validation of measurement based operations and validation of improving the output of the quantum algorithm by reducing qubit noise.

\subsection{Verification of functionality}
The first test evaluates the compiler's ability to compile a small quantum circuit into hardware executable instructions. Specifically, we verify whether the compiler correctly identifies the qubits that are being operated on and therefore decomposes operations properly. The compiler in this case is also verified to correctly identify the need for the electron qubit states to remain preserved, opting for the DDrf gate instead of direct control and therefore not breaking the state of the electron, resulting in faulty decomposition of the quantum circuit.

\begin{figure*}
    \centering
    \includegraphics[width=1\linewidth]{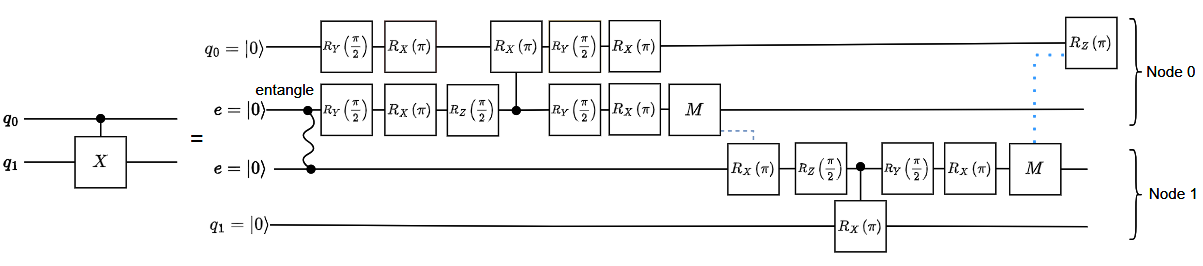}
    \caption{\textit{A full decomposition of a controlled X gate that is performed between two qubits that are on distant nodes (adapted from \cite{cx_gate}).
    The algorithm presented is a Controlled X teleportation algorithm in the native gateset of diamond NV centers. The left hand side shows the wanted operation, while the right hand side shows the operations needed to implement the operation in a diamond NV center}}
    \label{fig:teleportation_gate_algorithm}
\end{figure*}

\begin{figure}[]
    \centering
    \includegraphics[width=0.7\linewidth]{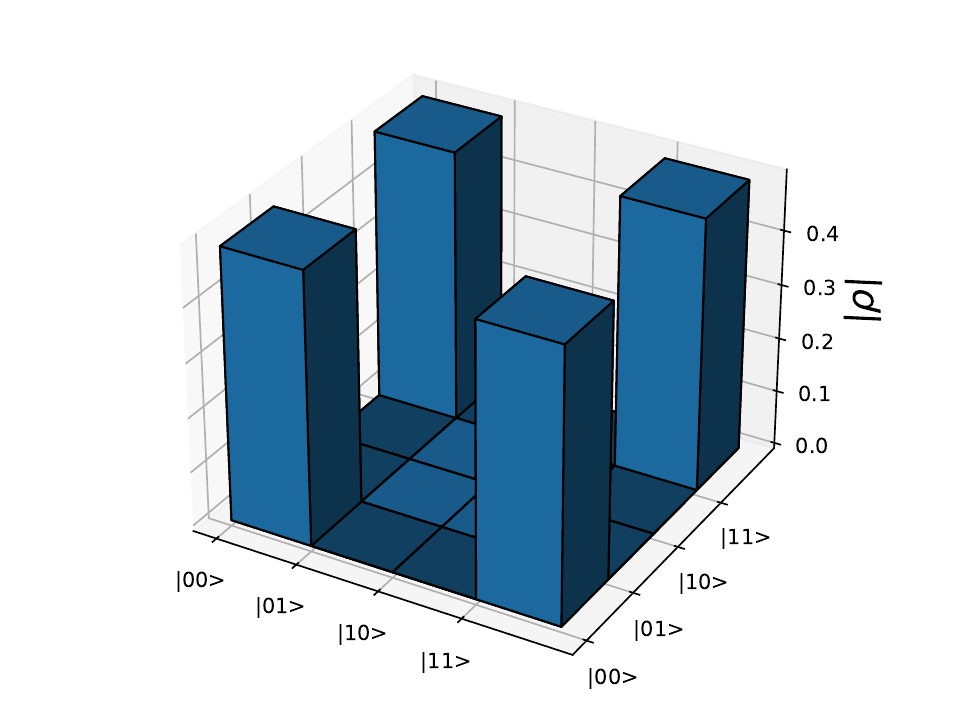}
    \caption{\textit{The output state of the simulator for the combined qubit state of two carbon qubits after a CNOT teleportation gate is performed on them. The control and target qubit were initialized in the $\ket{+}$, $\ket{0}$ state respectively. The presented state represents the $\frac{1}{\sqrt{2}}(\ket{00}+\ket{11})$ state.}}
    \label{fig:teleportation_gate}
\end{figure}

The selected test circuit is a teleportation-based CNOT gate between two qubits on different nodes. 
The algorithm implemented is depicted in Fig. \ref{fig:teleportation_gate_algorithm}. The control carbon qubit is initialized in the $\ket{+}$ state, while the target carbon qubit is initialized in the $\ket{0}$ state. The expected outcome of applying the CNOT gate to these qubits is the entangled state $\frac{1}{\sqrt{2}}(\ket{00} + \ket{11})$. The simulation output, presented in Fig. \ref{fig:teleportation_gate}, shows the resulting density matrix, which matches the expected outcome, confirming that the compiler correctly decomposed the distributed CNOT gate. This means that the compiler has correctly identified the need for the DDrf gate over the direct carbon gate and has performed proper routing for initialization of the carbon qubits. We can conclude this because a direct carbon gate would have destroyed the state of the electron, which would have resulted in a faulty output of the quantum circuit. 

% Given that the compiler has correctly decomposed the CNOT gate, this means that the compiler has correctly identified the need for a DDrf gate instead of a direct control gate for the carbon qubits. Also the compiler has correctly identified the qubits on which the operations are performed, rewriting the instructions into the qubit type specific instructions needed by the hardware.
% Existing compilers as if now only optimize for the theoretical implementation of the quantum circuit, and not for the hardware specific decompositions or qubit types.

\subsection{Verification of Measurement based operations}
The second test evaluates the compiler's capability to implement classically controlled operations. Specifically, we perform a measurement-based X gate on a carbon qubit, conditioned on the measurement outcome of an electron qubit. The carbon qubit is initialized in the $\ket{0}$ state, and the electron qubit is initialized in either the $\ket{0}$ or $\ket{1}$ state. 

\begin{figure}[]
    \centering
    \includegraphics[width=0.5\linewidth]{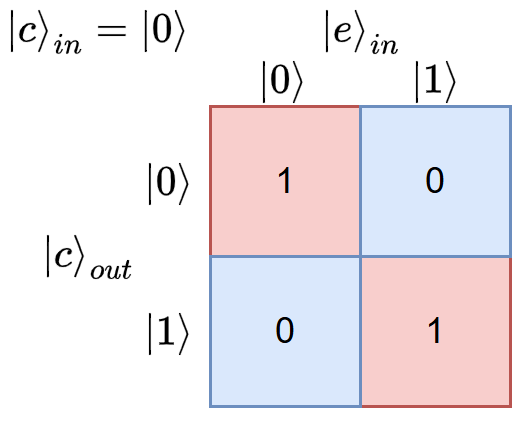}
    \caption{\textit{Confusion matrix
    measurement result for measurement based instructions, where the red and blue shaded areas represent that the measurement outcome happened or did not happen respectively. The carbon qubit is initialized in the $\ket{0}$ state, and the electron is either initialized in the $\ket{0}$ or $\ket{1}$ state. The results present that the output carbon state flips dependent on the state of the electron.}}
    \label{fig:measurement_based}
\end{figure}

The expected outcome of the measurement-based operation depends on the electron's state, yielding $\ket{0}$ or $\ket{1}$ for the electron states $\ket{0}$ and $\ket{1}$, respectively. The confusion matrix, shown in Fig. \ref{fig:measurement_based}, aligns with the expected values. In this figure, red shades indicate that a measurement outcome happened often, while blue shades indicate that the measurement outcome did not happen. These results confirm that the compiler successfully integrates classical and quantum instructions, enabling conditional operations essential for future quantum algorithms such as quantum error correction codes \cite{stabilizer_surface_codes}.% and variational algorithms \cite{}.

% This means that the compiler is capable of performing a measurement based operation, making use of the integration of classical and quantum instructions. Having the potential to perform measurement based operations is important for any future algorithms such as quantum error correction codes, some variational algorithms).

\subsection{Diamond specific optimization verification}
In the diamond NV center we can make use of direct carbon control and partial swaps to potentially reduce the total amount of gates and execution time of a quantum circuit. To evaluate these optimizations, we test whether the compiler's decompositions improve circuit fidelity in a noisy environment.

% Therefore, we have created a set of tests that verify that the decompositions given by the compiler actually result in improved circuit fidelity. The output of the compiler in the tests is simulated in a case with noise. 

\begin{figure}
    \centering
    \includegraphics[width=1\linewidth]{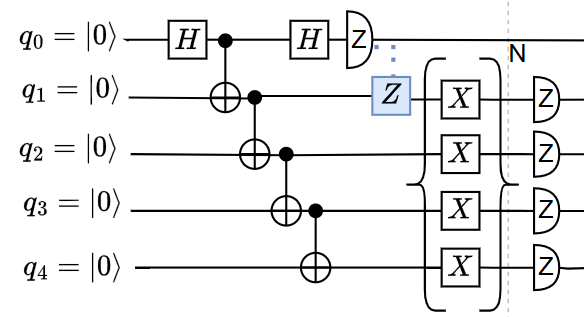}
    \caption{\textit{The quantum circuit used to test the influence of the compiler decompositions of the partial swap and direct control. The circuit is used to create a 4 qubit GHZ state only on the carbon qubits (1-4). The blue Z gate is a conditional gate that is only executed when the measurement value of qubit 0 is -1 (corresponding to $\ket{1}$).}}
    \label{fig:depol_decoh_test}
\end{figure}

We compare standard and optimized implementations of partial swaps and single qubit carbon gates by generating a 4-qubit local GHZ state using an NV center. The circuit is depicted in Figure \ref{fig:depol_decoh_test}, where for this first test, the physical X gates are not performed. The 4-qubit local GHZ state can represent a logical qubit encoded in the logical $\ket{0}$ state across four physical qubits. A key use case for partial swaps arises when performing consecutive measurements. This is because after a single measurement in a diamond NV center, the electron qubit state does not hold any useful information anymore. Therefore, any consecutive carbon measurement can make use of partial swaps. In our test, we create a 4 qubit initialized logical qubit and measure every qubit, where either a full or partial swap can be applied before performing the measurement. The partial swap is expected to reduce gate count while preserving correct outcomes.
 
The output of the compiler is first tested in the noiseless case to verify the functionality and afterwards the output is simulated using a noise model. In the latter we introduce depolarization noise (ranging from $0$ to $10^{-3}$), where the value represents the probability of depolarization happening on the qubit during a quantum instruction. This value is chosen because current diamond NV centers experience around this value for noise) and decoherence noise (coherence times ranging from $0.1$ to $100$ seconds, where the value represents the average time that a qubit remains in a coherent state. This noise represents outside noise happening on qubits that are in idle mode. This value is also chosen because current NV centers experience this value of noise \cite{coherence_time}.) during the operations to assess the performance under noise. 

% ?? why get rid of brackets?

In the noiseless case, the outcome of the logical measurement is calculated by multiplying all measurement outcomes of the physical qubits. The expected outcome is $1$ (representing $\ket{0}$, $-1$ represents $\ket{1}$) because the logical qubit is initialized in the $\ket{0}$ state. Our noiseless simulations also showed these expected results.
% The noisefull simulations 
In order to achieve reliable results for this test, the test must be performed multiple times, known as state tomography \cite{state_tomography}. In this case, we repeated the experiment $1,000$ times. Repeating the operations requires classical control, which is automatically added by the compiler, as explained in Section \ref{sec:classical_instructions}.

\begin{figure}[H]
    \centering

    \includegraphics[width=1.1\linewidth]{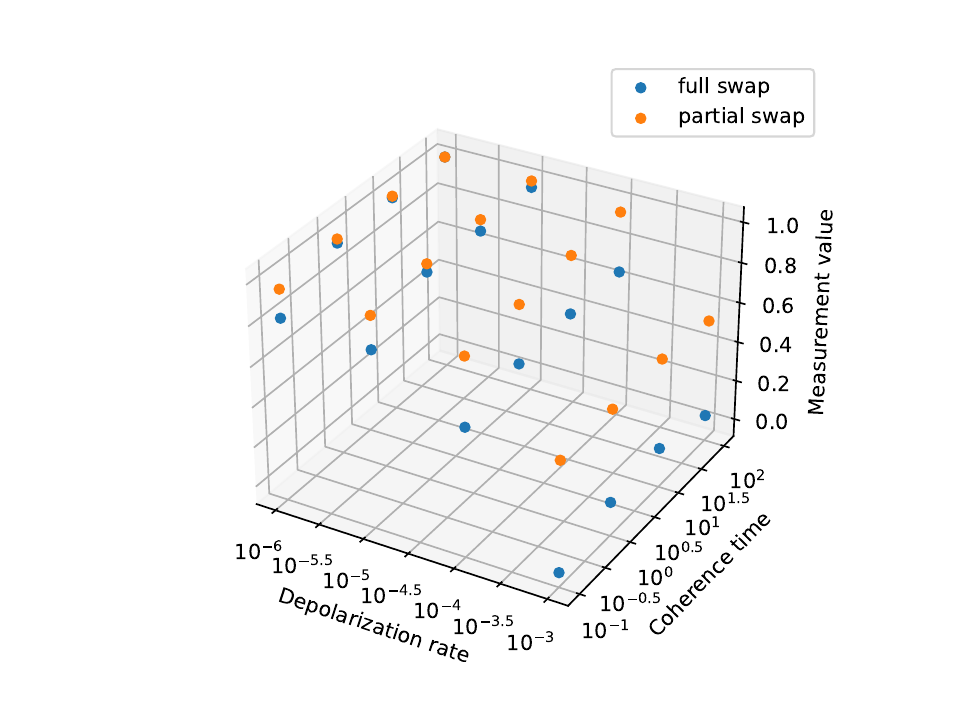}
    \caption{\textit{Depolarization and decoherence test for full swap versus partial swap mechanism. Four qubits are initialized in the logical $\ket{0}$ state and measured afterwards. The results show that performing a full-swap is more influenced by noise than the partial swap.}}
    \label{fig:swap}
\end{figure}

The results, presented in Fig. \ref{fig:swap}, show the average measurement value for the experiment (z axis) while sweeping both the depolarization (x axis) and decoherence (y axis) noise with noise values increasing. The orange data points show the results from the partial swap, while the blue marker show results from the full-swap. Since the logical state is $\ket{0}$, we expect the average measurement result to be close to $1$, therefore the closer the value on the z axis is to $1$, the better. In the figure, the results indicate that the full-swap operation is more sensitive to noise compared to the partial-swap, both in the depolarization and the decoherence noise regime. The results also show that the larger the noise value, the larger the difference between the full and partial-swap operation.

As the qubit count increases, the performance difference between the two swap methods is expected to become more pronounced, highlighting the advantage of partial swaps in reducing noise sensitivity. In the current test we have used a logical qubit encoded as $\ket{0}_L = \frac{1}{\sqrt{2}}(\ket{0000}+\ket{1111})$ and $\ket{1}_L = \frac{1}{\sqrt{2}}(\ket{0101}+\ket{1010})$. A logically encoded qubit with a larger amount of qubits (for instance 8 qubits) could  look like $\ket{0}_L = \frac{1}{\sqrt{2}}(\ket{00000000}+\ket{11111111})$ and $\ket{1}_L = \frac{1}{\sqrt{2}}(\ket{01010101}+\ket{10101010})$, where reading out the qubit would result in 8 physical qubit measurements. Performing partial swaps instead of full swaps to perform these measurements would allow for even more reduction of noise effects, compared to the 4 qubit logical encoded state.

This result is crucial for future quantum algorithms involving logical qubits, which will be mapped onto multiple physical qubits. This is the basis for future algorithms, after the NISQ era. For instance, quantum error correction codes make use of logical encoded qubits. In such cases, the partial swap mechanism is particularly useful, as the electron qubit typically serves as a communication qubit and its state is often not important when performing logical operations. Moreover, in the future, the logical qubit might even be mapped to more physical qubits, further improving the use of the partial swaps.

\begin{figure}[H]
    \centering
    \includegraphics[width=1.1\linewidth]{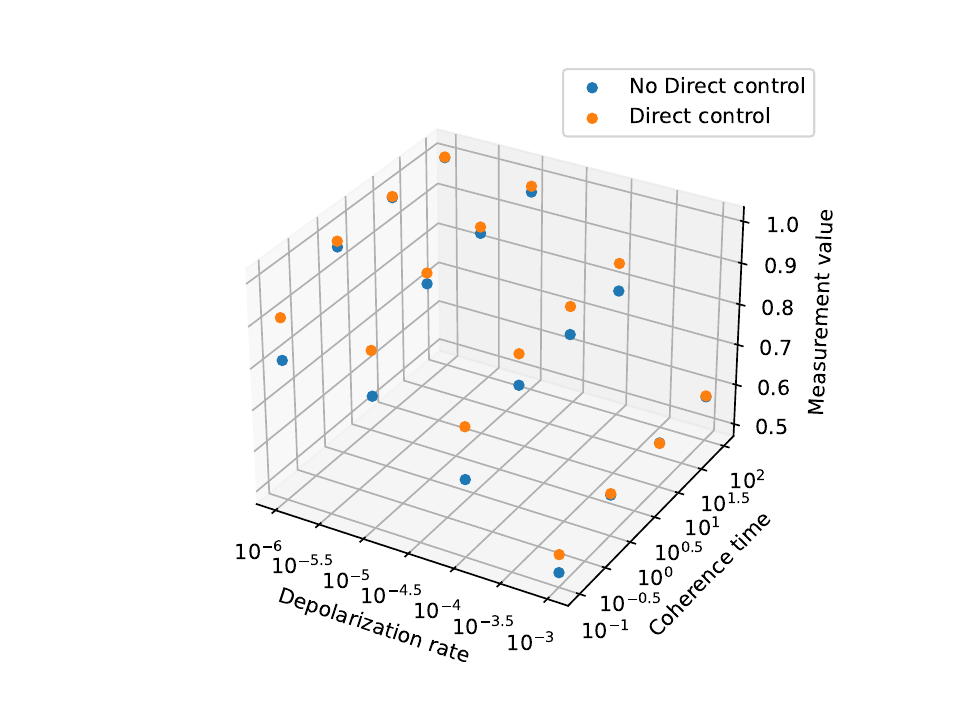}
    \caption{\textit{Depolarization and decoherence test for direct versus indirect carbon control mechanism. Four qubits are initialized in the logical $\ket{0}$ state and an even amount of X gates is performed on them. The results show that performing indirect control is more influenced by noise than the directed control.}}
    \label{fig:preserve}
\end{figure}

Additionally, we compare direct and indirect carbon control by generating the 4 qubit local GHZ state, but now we perform multiple (and an even amount of) physical X gates on the qubits. In this case the expected outcome is again $1$, because after an even number of X gates the logical qubit is back at the $\ket{0}$ state. These X gates are decomposed into instructions using either direct or indirect control of the carbon qubits. To evaluate the effect of direct control, the same noise sources as for the partial swap test are introduced during the operations. In this case we expect the direct control to result in lower sensitivity to noise because the direct control uses fewer gates (reducing the depolarization noise) and takes a shorter amount of time (reducing the decoherence noise). The results, depicted in Fig. \ref{fig:preserve}, show the average measurement value for the experiment (z axis) while sweeping both the depolarization (x axis) and decoherence (y axis) noise with increasingly noisy values. The orange data points show the results from the indirect control implementation, while the blue marker represent results from the direct control implementation. The measurement values in the noiseless case are expected to be 1, therefore the closer the measurement data (z axis) is to 1, the beter the result. The results demonstrate that when direct control is applicable, it reduces the noise influence both due to depolarization as due to decoherence, leading to improved performance.

This demonstrates the value of direct carbon control in future decompositions from logical to physical qubits, because a logical instruction will in most cases decompose into multiple physical instructions, where the electron state does not need to be preserved. The amount of physical instructions to represent a logical instruction scales with the size of the logical qubit. If we again look at the logical 8 qubit encoded state presented before, one can see that a logical X gate would require twice as much physical X gates to transform from the logical 0 to the logical 1 state with respect to the 4 qubit logically encoded case. Therefore, these decompositions would benefit from using direct control where the benefit scales with the size of the logical qubit.

% ?? reads more as eval without discussion. How to solve? 

%% file: Chapters/Conclusion.tex
\section{Conclusion} \label{sec:conclusion}
 
In previous works, general compilers have been presented capable of rewriting quantum instructions into standard gate sets, as well as compilers tailored to specific technologies like superconducting qubits. However, in this work we have presented the first diamond NV center specific compiler. In addressing our research question, we proposed new compiler passes and introduced diamond-specific operations and optimizations.
Our compiler performs diamond NV center-specific decompositions and optimizations, reducing noise effects and provides full control over both quantum and classical instructions. Classical instruction control enables system diagnostics and measurement-based operations with a high degree of integration.

% In previous works general compilers have been presented capable of rewriting quantum instructions into standard gate sets, as well as compilers tailored to specific technologies like superconducting qubits. However, in this work we have presented the first diamond NV center specific compiler. Our compiler performs diamond NV center-specific decompositions and optimizations, reducing noise effects and provide full control over both quantum and classical instructions. Classical instruction control enables system diagnostics and measurement-based operations with a high degree of integration.

The main features of our compiler include support for diamond specific operations such as direct carbon control and partial swaps, useful for carbon qubit initialization, measurement, or mid-circuit swaps whenever the electron qubit state can be discarded. The compiler also supports classically controlled operations, state tomography and system diagnostics. 

To evaluate performance and noise mitigation, we compiled a CNOT-based teleportation algorithm and simulated its execution. The results confirm the generation of hardware-executable, NV-compatible instructions that effectively incorporate measurement-based operations.
% Our compiler has been evaluated for functionality and influence on noise mitigation by decomposing a CNOT teleportation algorithm and running the output on a simulator.
% The results demonstrate that the compiler can generate hardware executable diamond-based quantum instructions suitable for the target hardware, utilizing diamond specific operations. The algorithm uses measurement based operations, showing that the compiler can create instructions that represent measurement based operations.
% The compiler has been used to successfully implement a CNOT teleportation algorithm, decomposing the quantum circuit into hardware executable instructions, where the output is tested using a micro-architecture simulator that outputs the results of the circuit. 
% The output of the simulator matched the expected results, verifying that the compiler output is correct. Within this algorithm, there was also a measurement based operation, therefore this also shows that the compiler is capable of creating a sequence of instructions implementing a measurement based operation.  

Classically controlled operations are a vital part of certain algorithms such as quantum error correction codes \cite{stabilizer_surface_codes}. These types of algorithms are particularly important, because they are needed to progress quantum computers out of the NISQ era.

We further tested the compiler's diamond-specific features by generating a 4-qubit GHZ state, applying X gates, and comparing measurement outcomes using both full and partial swaps. Simulations under depolarizing and decoherence noise models showed improved fidelity when using partial swaps and direct carbon control versus full swaps and indirect carbon control.  When logical qubit sizes grow, the amount of physical operations needed to perform a logical operation also grows. Therefore, these features are expected to scale with the logical qubit size, offering increasing benefits in larger systems. The compiler also automatically integrates state tomography for result verification.

Future work includes extending this compiler framework to other quantum hardware platforms that could benefit from automatic quantum operation optimization. Additionally, further optimizations for NV center-based quantum computing could be explored, such as multi-CNOT gate control across multiple NV centers or improved logical-to-physical decompositions to minimize photon entanglement requirements. These advancements would further enhance the compiler’s ability to optimize quantum circuits. At last, automatic Pauli string measurement or system diagnostics for other technologies can be added.

%% file: Chapters/acknowledgement.tex
\section{Acknowledgments}
We gratefully acknowledge support from the joint research program “Modular quantum computers” by Fujitsu Limited and Delft University of Technology, co-funded by the Netherlands Enterprise Agency under project number PPS2007.